# Low charge noise quantum dots with industrial CMOS manufacturing

A. Elsayed[1, 2], M.M.K. Shehata[1, 2], C. Godfrin[1], S. Kubicek[1], S. Massar[1], Y. Canvel[1], J. Jussot[1], G. Simion[1], M. Mongillo[1], D. Wan[1], B. Govoreanu[1], I. P. Radu[1], R. Li*[1], P. Van Dorpe[1, 2], K. De Greve[1, 3]

[1] IMEC, Leuven, Belgium
[2] Department of Physics and Astronomy, KU Leuven, Leuven, Belgium
[3] Department of Electrical Engineering, KU Leuven, Leuven, Belgium

Silicon spin qubits are among the most promising candidates for large scale quantum computers, due to their excellent coherence and compatibility with CMOS technology for upscaling. Advanced industrial CMOS process flows allow wafer-scale uniformity and high device yield, but off the shelf transistor processes cannot be directly transferred to qubit structures due to the different designs and operation conditions. To therefore leverage the know-how of the micro-electronics industry, we customize a 300mm wafer fabrication line for silicon MOS qubit integration. With careful optimization and engineering of the MOS gate stack, we report stable and uniform quantum dot operation at the Si/SiOx interface at milli-Kelvin temperature. We extract the charge noise in different devices and under various operation conditions, demonstrating a record-low average noise level of 0.61 $\mu eV/\sqrt{Hz}$ at 1 Hz and even below 0.1 $\mu eV/\sqrt{Hz}$ for some devices and operating conditions. By statistical analysis of the charge noise with different operation and device parameters, we show that the noise source can indeed be well described by a two-level fluctuator model. This reproducible low noise level, in combination with uniform operation of our quantum dots, marks CMOS manufactured MOS spin qubits as a mature and highly scalable platform for high fidelity qubits.

## Introduction

The demand for greater computational power has put quantum computation in the limelight: the concept of using quantum states for information processing promises a huge speedup and solutions to certain problems that are intractable on classical computers[1]. In the past decades, quantum computing has shown marked development, with small-scale quantum algorithms and quantum advantage achieved on different platforms[2–4]. In moving towards practical quantum computing applications, the research focus is shifting from fundamental qubit operation to large qubit systems[5]. Particularly, qubits fabricated with industrial semiconductor manufacturing technologies recently received great attention[6–8]. The capability of the semiconductor industry in fabricating billions of nanostructures with remarkable uniformity and reliability could indeed be leveraged for the full integration of large-scale quantum processors[9,10]. The rationale here is that the first practical quantum algorithm would require the number of physical qubits to be in the million scale[11], similar to the transistor count of integrated circuit chip transistors back in 1990s[12].

Tantalizingly, the structure of silicon spin qubits closely resembles the CMOS transistor technology[13]. Nanoscale electrodes define the quantum dot structure trapping a single spin of an electron or a hole[14]. With industrial fabrication, semiconductor nanostructures are patterned subtractively for accurate wafer-scale critical dimension control, where lithography defined patterns are transferred onto the gate electrodes by dry etching rather than the laboratory-based liftoff process[15]. Comprehensive metrology tools and close process monitoring steps ensure high reproducibility and yield. Moreover, the material science and characterization technology developed around the CMOS industry allow deep insight into the performance limiting factors and optimization directions. All this could provide spin qubits a shortcut for upscaling.

The spin state is also an excellent platform for quantum information encoding and processing[16]. By isotopically purifying the silicon substrate, long quantum coherence has been achieved[17,18]. Single and two qubit gates with operation fidelities higher than the error correction threshold have been demonstrated and efforts towards scaling have been shown with a six-qubit device[17,19–23]. In addition, spin qubits can be operated at elevated temperatures above 1 K, addressing several upscaling requirements on cooling power and wiring interconnect[24,25]. For the next step towards large qubit arrays, advanced semiconductor manufacturing is commonly expected to be needed to realize the architectures required for scaling up[16,26,27]. However, state-of-the-art spin qubits are mostly fabricated in laboratory-based environments[17–23]. Though there are several exciting demonstrations of qubits made by advanced industrial fabrication with good yield and transport uniformity, the final qubit performance typically shows a certain level of degradation over that of lab devices[6–8]. Device charge noise, which is one of the limiting factors for spin coherence and an important metric characterizing the device quality at cryogenic temperatures[19,28–32], is typically high in fab structures[8,33,34].

To minimize the charge noise, it is important to understand its origin. The semiconductor/oxide interface has been identified as the source of charge noise and has been intensively studied over the last 40 years[35]. The McWhorter model is widely accepted in MOSFET devices[36]. In this model, defects in the oxide cause carrier number fluctuations in the MOSFET channel. However, in quantum devices it has been shown that charge noise is not a result of carrier number fluctuations. Rather, defects at the Si/SiOx interface or in the oxide, for these quantum devices, can be described as bistable systems known as two-level fluctuators (TLFs)[37]. Each TLF has a unique Lorentzian spectrum. A random distribution of TLFs, all added together, results in the infamous $1/f$ noise[38]. However, the microscopic nature of the TLFs remains inconclusive. While many different proposals have been employed to explain the origin of the TLFs including tunneling atoms or tunneling electrons[39,40] it remains difficult to identify microscopic defects from typical measurements. Low noise devices with high quality interfaces could reveal individual TLFs and show their microscopic origin.

In this work, we customize a state-of-the-art 300mm wafer fabrication flow for silicon MOS qubit structures[7]. Through full gate stack optimization, we demonstrate that the Si/SiOx interface could provide low-noise environments for qubit operations rather than detrimental. Across multiple devices, all the quantum dot structures show stable and uniform operation at milli-Kelvin, and an average charge noise level of 0.6 μeV/$\sqrt{Hz}$ at 1 Hz. This ultra-low noise level at different devices and operation conditions allows statistical analysis of the key metrics of the charge noise spectrum. With numerical simulations, we find that the charge noise can be well described with a simple TLF model and provide further insights into spatial density and position.

## Device fabrication and interface characterization

Industrial CMOS fabrication processes are used for our device fabrication[7]. We optimize the process flow for spin qubits, specifically for the planar overlapping gate scheme[41]. The overlapping structure has been the main design for high fidelity qubit operations[17,20,22,23], and the integration with the standard industrial fabrication allows wafer-scale accurate critical dimension control and high device yield[7,8,10,33]. We also employ 300mm e-beam lithography for the pitch critical quantum dot gates and optical lithography for the size relaxed features. E-beam enables fast turnaround for device designs, and the process can be transferred to advanced optical lithography as similar photoresist and hardmask are used. The fabrication starts with a 12 nm thermally grown oxide, which defines the high-quality Si/SiOx interface for quantum dot confinement along the vertical direction. Following that, we deposit the first gate layer across the wafer and pattern it subtractively. Rather than metallic gates, we use polysilicon as the gate metal to reduce interface strain at cryogenic temperatures[33,42]. Polysilicon also allows the integration of a high temperature oxide as the inter-gate dielectric as opposed to depositing low temperature dielectrics which typically host additional defects[43]. By repeating the above processes, we achieve the overlapping gates as shown in Fig. 1a.

However, subtractive patterning with dry etching can be more intrusive than laboratory-based metal lift-off processes[44]. This is further exacerbated for overlapping gate devices, in which each device is subjected to multiple etch steps[33]. The degraded dielectric and interface contribute to quantum dot non-uniformity and higher charge noise. To overcome this, we carefully optimize the process flow, and study the Si/SiOx interface with cryogenic hall mobilities[45], as shown in Fig. 1b (methods). In comparison to the first gate level, we find that the mobility drops considerably for higher gate layers. Nonetheless, the degradation is significantly reduced with optimized etching conditions, guaranteeing high quality structures for full qubit integration.

We perform further quantum transport measurements on the first gate layer, where the Si/SiOx is protected by the gate from the following processing. The quality of the primary Si/SiOx interface can be evaluated for a large-scale qubit array as well as for the implementation of the single-etch gating scheme[46]. With a base temperature lower than 10 mK, we can reach a peak mobility of $30 \times 10^3$ cm$^2$/Vs at a charge density of $4 \times 10^{11}$ cm$^{-2}$ (Fig. 1c), which is significantly higher than previously reported Si MOS Hallbars with ~ 10 nm SiOx gate oxide[45,47]. From a Metal-to-Insulator fit[48] we determine a percolation density of $8 \times 10^{10}$ cm$^{-2}$. This matches the lowest values reported for MOS gate stacks[45,47] and is comparable to SiGe heterostructures[49]. To gain further insight into the interface quality limiting factor, we operate the device in the quantum hall regime as shown in Fig. 1d and extract the Dingle ratio[50] (see Supplementary Information section 1). The Dingle ratio is an indication of the spacing between the scattering centers to the conduction channel in 2D electron gas (2DEG) systems, with higher values mean larger spacing. We report a Dingle ratio ~3, which is not as high as GaAs or SiGe heterointerfaces[51,52] but larger than the typically number of ~1 on Si MOS samples[45], suggesting that rather than defects directly at the Si/SiOx interface, defects in the oxide also play an important role in our optimized gate stack.

## Electrical characterization of single quantum dots

For spin qubits, gate defined quantum dots are used to trap single spins[14]. Beyond the single spin regime, the quantum dots can be used as charge sensors to read out the spin states via different spin-to-charge conversion methods[53,54]. Only a single charge can flow through the quantum dot at any given time due to Coulomb repulsion, and the current is very sensitive to the environmental electrical potential[55]. Such a structure is called a single electron transistor (SET).

As shown in Fig. 2a, the SET highly resembles a planar MOSFET transistor but with more gate electrodes[56]. The SET top gate (ST) induces an electron channel while the left and right barrier gates (LB and RB) define the single quantum dot in between, which is a nanoscale electron island. Fig. 2b shows the quantum dot charge stability map by measuring the SET current as a function of biases on LB and RB. The diagonal lines correspond to Coulomb oscillations where electrons tunnel through the quantum dot one by one[55]. Between the Coulomb oscillations, the electron numbers are fixed.

To study the SET uniformity, we measure the $I_D V_G$ characteristics by sweeping each gate individually while keeping the other two at a fixed high voltage potential. In Fig. 2c, we show the average $I_D V_G$ characteristics of 12 devices (12 ST gates and 24 barrier gates). Furthermore, we extract the threshold voltage and sub-threshold swing at standard maximum transconductance point (methods). The devices show remarkable uniformity as seen in the CDF plots in Fig. 2b, with a standard deviation ~ 70 mV for the barrier threshold voltage and < 20% for the sub-threshold swing. This highlights the uniformity with CMOS manufacturing and provides guidelines for the design of a large-scale spin qubit array[8].

# Charge noise

The electron spin has long coherence as there is no direct coupling to charge fluctuations. By isotopically purifying the silicon substrate, recent studies show single spin dephasing times T$_2$* beyond 10 μs and even 100 μs[17–19,28]. The remaining spin-orbit interaction (SOI), either intrinsic from the Si/SiOx interface[41] or extrinsic with micromagnet structure[57], indirectly couples the spin to the environment's electrical field noise. Many of the best reported spin qubits show that the device charge noise is in fact the limiting factor for the qubit coherence and the quantum gate fidelity[28,58].

To study the device noise, we focus on the SET charge noise as it has been shown to well represent the qubit noise , and typically the noise figure at 1 Hz is used as the metric to benchmark between different structures and material platforms[28–30]. On the flank of the Coulomb peak, where the SET is the most sensitive to environmental noise, we record the current noise spectrum. We extract the quantum dot potential fluctuation from this current noise with the device transconductance and capacitance ratio (methods), and the final charge noise spectral density is shown in Fig. 3a. This measurement is repeated on all Coulomb peaks ranging from $V_{ST} = 3.5$ V to 4.0 V, as shown in Fig. 3b. For statistical analysis, we perform the same procedure across 12 different devices, which gives 231 different spectra in total. In Fig. 3c and d, we show violin plots of the charge noise and the power factor extracted from each noise measurement, respectively (see Supplementary Information section 3). The charge noise shows a standard deviation of $0.27\,\mu eV/\sqrt{Hz}$ with the average of $0.61\,\mu eV/\sqrt{Hz}$ and several points lower than $0.1\,\mu eV/\sqrt{Hz}$. This is the state-of-the-art low charge noise for MOS devices[8,32–34], and comparable to SiGe heterostructures[28,29,31,51].

On closer examination of each charge noise spectrum, we found that most have a power law dependence $1/f^\alpha$, where $\alpha$ is the power factor. This indicates a wide spatial distribution of TLFs but different frequency distributions as $\alpha$ deviated from 1 slightly in some cases, which has also been recently observed in Si/SiGe quantum dots[31]. However, some spectra are Lorentzian in type[34], or even the combination of a power law and Lorentzian, as seen in Fig. 4a. The Lorentzian noise spectrum suggests the presence of a single dominating TLF[38], which can be expected given the random site of TLFs and the nanoscale quantum dots; for sufficiently low average densities of TLFs, statistical fluctuations in nanoscale devices are expected to result in some devices only observing a (few) dominant TLF(s).

The nanoscale quantum dots together with the low charge noise spectra could therefore serve as an excellent probe for TLFs and their microscopic nature. In Fig. 4b, we plot each $S_0$ at the corresponding ST bias. Across an ST bias range from 3.5 to 4 V, we see a uniform distribution of $S_0$. In other words, as we change $V_{ST}$, different ensembles of TLFs are activated or deactivated randomly, but the overall active number of TLFs at different electrical field is the same. We further examine the charge noise with respect to different metrics (see Supplementary Information section 3). No correlation between $S_0$ and $\alpha$ is observed, suggesting that the TLFs are uniformly distributed in frequency due to the wide spread of the relaxation times. Meanwhile, we see a wide distribution of $\alpha$ between 0 and 2, suggesting the spectra dataset has a good coverage of TLFs with different signature frequencies. In addition, varying the ST bias changes the number of electrons in the quantum dot, and we detect a systematic change in the quantum dot size and coupling ratio of the dot to ST. As electrons in the quantum dot could partially screen the electrical field, varying the electron numbers could influence the charge noise pickup, especially in the few-electron regime[31,59,60]. But we do not identify a correlation between the charge noise and the lever arm, which is the ratio of the ST-to-dot capacitance to the overall dot capacitance, suggesting that the screening effect is weak or change little in our SET structures. The above results suggest a uniform distribution of TLFs, which agrees with the standard TLF model[38]. Similar results have been reported with superconducting qubits[35], where most results are limited to fast TLFs in the radio frequency regime due to the nature of the microwave resonators in the qubits acting as a frequency filter. Our results suggest that near DC, i.e., nine frequency decades away, slow TLFs around 1 Hz are also uniformly distributed at the Si/SiOx interface and/or in the oxide, as shown schematically in Fig. 4c.

## Noise simulations

Our statistical measurements and analysis on the quantum dot charge noise suggest its microscopic origin: randomly distributed TLFs over a nanoscale device area. To further verify this postulation, we employ a simplified TLF model to reconstruct and validate the experiments. Following our previous study in ref 61, we use double well potential (DWP, charge dipole) type defects to represent the TLFs as they best match the quantum dot charge noise, since the whole electron trapping / detrapping would lead to much larger noise distribution than what we observe in our data – as well as much larger than in previously published literature results[8,31]. We assume a two-dimensional distribution of TLF in the silicon oxide, and only the TLF density ($n_{TLF}$) and its depth in the oxide ($z$) are tunable variables in the simulation. With the above conditions, TLFs are randomly placed over an area centered around a many electron quantum dot. We perform a Monte Carlo simulation with 1000 random sets of TLF configurations and extract each noise spectrum of the quantum dot potential. Further details about the simulation can be found in Supplementary Information section 6.

Fig. 5a shows the average noise at 1 Hz ($\bar{S}_0$) as a function of $n_{TLF}$ and $z$. In Fig. 5 b and c, we plot the kernel density for the $S_0$ and $\alpha$. Decreasing $n_{TLF}$ or reducing $z$ increases the distribution of $S_0$ and $\alpha$ around the averaged values, as the probability of finding a single dominating TLF increases. We find $n_{TLF} = 1.6 \times 10^{10}$ cm$^{-2}$ and $z = 6$ nm to best represent the experimental distributions as shown in the overlaying distributions in Fig. 5d and e. The good simulation-to-experiment agreement on charge noise and power factor for both average values and distribution further supports that randomly distributed TLFs over the quantum dot lead to the device charge noise. In addition, we would like to point out that other combinations of $n_{TLF}$ and $z$ could still have reasonable match between simulation and experiments, and three-dimensional distribution of TLF could give a better matching than simple two-dimensional (Supplementary information section 6). Nonetheless, this simulation suggests that disorders at the Si/SiOx interface does not need to be detrimental to quantum dot performance, and that deep in the oxides or even at the upper oxide/gate interface should be also taken into consideration for further optimization[62].

This simple TLF model also enables the construction of a charge noise environment to predict the qubit performance. With the co-simulation framework reported in ref 61, we estimate the two-qubit SWAP gate fidelities with above TLF distribution (see Supplementary Information section 7). The average SWAP error in our models is lower than 0.1%, highlighting the low noise level of Si MOS quantum dots with industrial manufacturing.

## Conclusion

We show the integration of Si MOS quantum dots with full industrial 300 mm wafer fabrication. By leveraging the know-how of CMOS technology, we optimize the full gate stack for high performance spin qubit devices with high yield and uniformity. The Si/SiOx interface shows state-of-the art Hall mobility and critical density. Considering the nanoscale nature of the quantum dots versus the full 300 mm wafer, we characterize multiple SETs for statistical analysis. The SET gates demonstrate uniform turn on curves and single quantum dot formation in the milli-Kelvin temperature range. We further characterize the SETs using current spectroscopy techniques and extract record-low average charge noise of 0.61 μeV/√Hz on Si MOS structures. The stable, uniform, and low noise operation is comparable to best reported SiGe heterostructure base devices, which shift the interface to the deeper Si/SiGe quantum wells to reduce the influence from disorders. However, the Si/SiGe interface brings significant drawbacks in terms of valley splitting[47,63] as well as crosstalk[64], limiting the yield rate and addressability of large qubit arrays. Our results confirm the Si MOS is and remains a compelling material platform for spin qubits and the maturity of industrial fabrication for the qubit development.

The low noise quantum dot together with its nanoscale size could be an excellent probe for TLFs and their microscopic nature. The statistical analysis of the charge noise shows uniform distributions with respect to different metrics. Noise spectra reconstructed with a simple TLF model can also match the experiments well. These findings are strong evidence that the quantum dot charge noise indeed originates from randomly distributed DWP type defects over the nanoscale area. Future studies could investigate the influence of gate oxide thickness on charge noise and explore novel detection methods for slow TLFs across the gate stack.

## Methods

**Hall bar characterization.** The fast turn-over interface characterization is carried out in a cryogenic probe station with base temperature of 4 K and a 2.5 T magnet. This allows us to refine the process and preselect high quality devices with optimized processes for further characterization. The advanced interface characterization is carried out in a dilution fridge with a base temperature of 8 mK and a 3 T magnet. Both setups employ a standard Hall bar measurement procedure where a positive voltage is applied to the top gate to form a 2-dimensional electron gas (2DEG) at the Si/SiO$_X$ interface. The current through the sample is measured through an IV transimpedance amplifier. Additionally, the longitudinal and transverse Hall voltages are measured simultaneously using standard lock-in techniques. The Hall carrier sheet density $n_S$ is extracted from the Hall resistance $R_{xy} = B \cdot n_S \cdot e$, where $B$ is the magnetic field, $n_S$ is the 2DEG density, and $e$ is the electron charge. The carrier mobility $\mu$ is obtained from $\mu = n_S \cdot e \cdot \rho_0$ where $\rho_0$ is the resistivity. The percolation density is extracted from a metal-to-insulator transition (MIT) fit of the density dependent conductivity $\sigma \sim (n - n_p)^{1.31}$. The measurement is repeated on three different Hall bars across the wafer, and we found good agreement.

**Quantum dot threshold voltage and subthreshold swing.** A fourth order routine is used to calculate the derivative of the measured $I_D V_G$ data. The maximum derivative (maximum transconductance $g_m$) of the curve is used to calculate the threshold voltage $V_{TH}$ and the conductance $K_0$. All gates of the quantum dot were operated in the linear regime where the drain current is given by $I_D = K_0(V_G - V_{TH} - V_D/2)V_D$. Thus, the $I_D V_G$ is used to fit $K_0$ and $V_{TH}$. The conductivity $K_0$ equals the fitted slope divided by the drain voltage. The threshold voltage $V_{TH}$ is the intercept of the tangent line $g_{m_{\max}}$ with the x-axis minus $V_D/2$.

**Electrical characterization of the quantum dots.** The devices are tuned systematically with the following procedure. From the barrier gate stability map, we identify the barrier gate voltage. We then sweep the top gate bias from 3.5 V to 4.0 V to measure the Coulomb oscillations. If a background current is observed between the Coulomb oscillations, the barrier gates are readjusted, and the measurement repeated. The source drain bias is then swept with respect to the top gate voltage to obtain Coulomb diamonds. We measure the drain current noise spectrum $S_I$ at the maximum sensitivity point on the left flank of each of the Coulomb peaks with 30 averages over 3 minutes The current noise spectrum is converted to charge noise spectrum using $S_0 = \alpha \sqrt{S_I}/(dI/dV_{ST})$, where $\alpha$ is the lever arm of the quantum dot, and $dI/dV_{ST}$ is the slope of the Coulomb peak. Further details on extracting the quantum dot charge noise can be found in the Supplementary Information section 2.

## Data availability

The data that support the findings of this study are available from the authors upon reasonable request.

## Code availability

The codes used in this study are available from the authors upon reasonable request.

## Acknowledgements

The authors acknowledge financial support from European Union's Horizon 2020 Research and Innovation Program under grant agreement No 951852 (QLSI). This work was performed as part of IMEC's Industrial Affiliation Program (IIAP) on Quantum Computing.

## Author contributions

A.E., M.M.K.S., and C.G. performed the experiments. S.K. integrated the devices with support from S.M., Y.C., and J.J.. M.M.K.S. performed the simulation with support from G.S.. R.L., M.M., D.W., B.G. I.R., P.V.D. and K.D.G. conceived and supervised the project. A.E., R.L., and K.D.G. wrote the manuscript with input from all the authors.

## Competing interests

The authors declare no competing interests.

# Figures

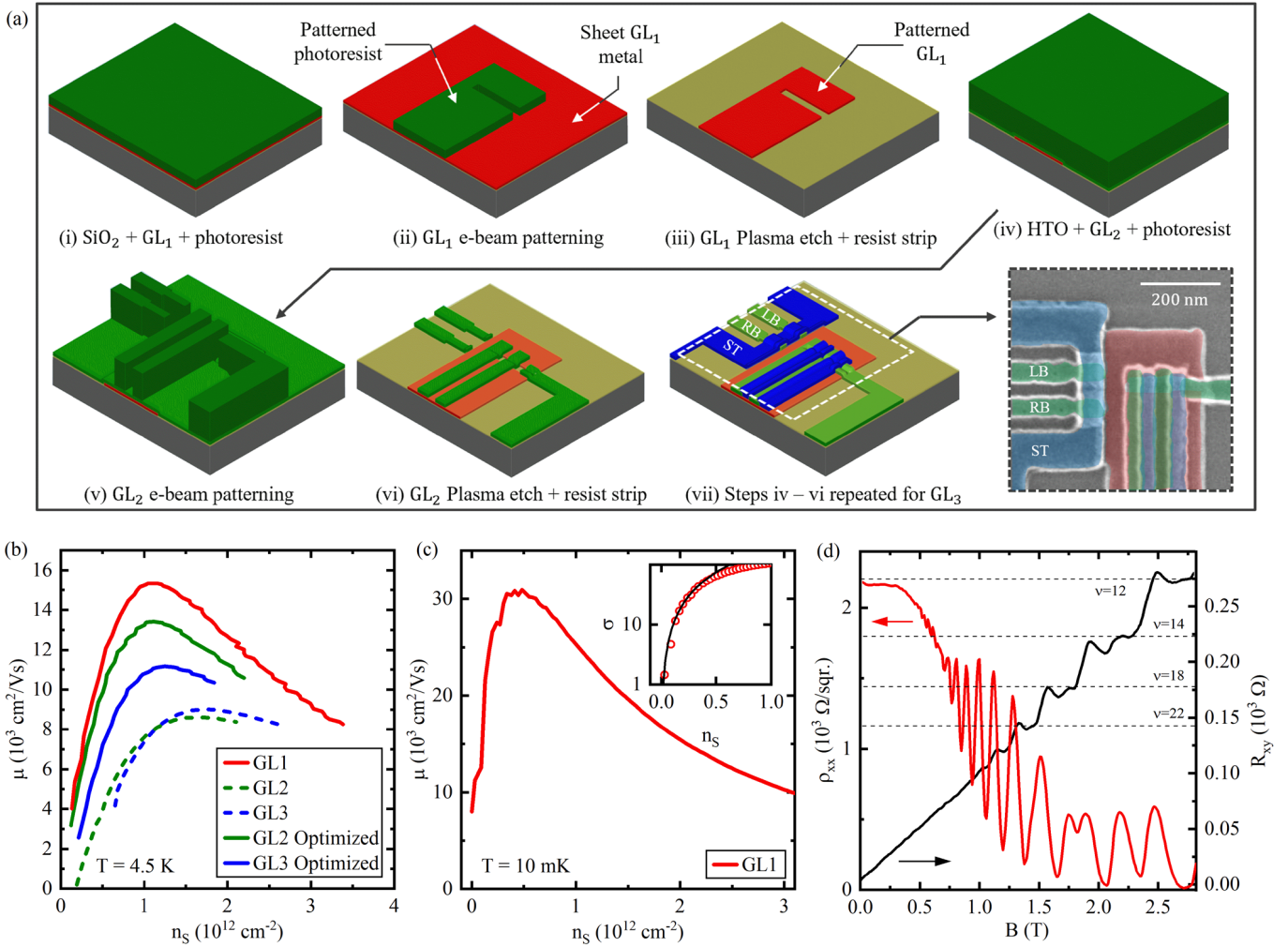

**Fig. 1| Device fabrication and interface characterization.** a) Schematic of the subtractive patterning steps to fabricate the overlapping qubit structures. (i) A full sheet of gate material and photoresist are deposited through the 300 mm wafer. (ii) The photoresist is patterned with electron beam lithography. (iii) The 1$^{st}$ gate layer is patterned using dry etching techniques. (iv) For the 2$^{nd}$ gate layer, the steps start with deposition of high temperature oxide (HTO), 2$^{nd}$ gate layer, and photoresist. (v) Then the photoresist is patterned with e-beam, and (vi) transferred to the gate material with dry etching. (vii) Step iv-vi are repeated for the 3$^{rd}$ gate layer. The resulting scanning electron microscope (SEM) image of the qubit structure is shown in the bottom right. b) Hall mobility of the Si/SiOx interface under different gate layers. The effect of multiple patterning steps is evident in the reduced mobility of higher gate layers (gate level 2 and 3, dashed lines). Through process optimization we show an improved mobility for gate level 2 and 3 (solid lines). c) Hall mobility of gate level 1 at 10 mK, showing a peak mobility of of $30 \times 10^3\ cm^2/Vs$ and a percolation density of $8 \times 10^{10}\ cm^{-2}$. d) Quantum Hall effect of gate level 1 showing clear plateaus in the transverse resistance ($R_{xy}$) and SdH oscillations in the longitudinal resistance ($\rho_{xx}$). The SdH oscillations are visible starting $B = 0.5\ T$ and the oscillation minima go to zero at $B = 2\ T$, further indicating a high quality single subband transport channel.

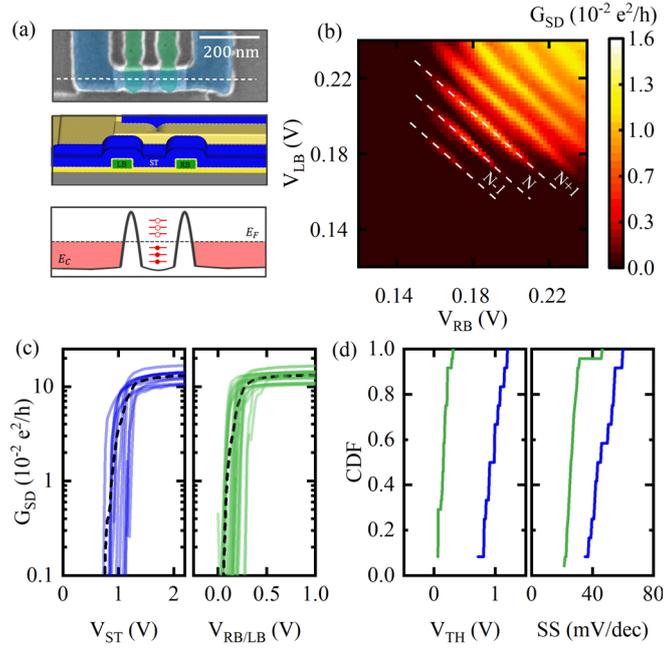

**Fig. 2| SET operation and device uniformity.** a) The SET device structure. The top is the SEM top view. At the white dashed line, the schematic cross section is shown in the middle. The bottom shows the band diagram, where the SET top gate (ST) forms the electron channel and the left and right barrier gates (LB and RB) induce the barriers above the Fermi level and define the quantum dot. b) Barrier gate stability map at a fixed $V_{ST} = 3.5 V$. The single quantum dot between the barriers gives the diagonal Coulomb oscillation lines in the map (white dashed lines). c) $I_D V_G$ characteristics of the top gate (blue lines) and barrier gates (green lines) across 12 SETs at 10 mK. d) The cumulative distribution functions (CDF) of the threshold voltage ($V_{TH}$) and subthreshold swing ($SS$) of the SET gates. The almost vertical lines, especially for the barrier gates, highlight the level of uniformity with advanced manufacturing techniques.

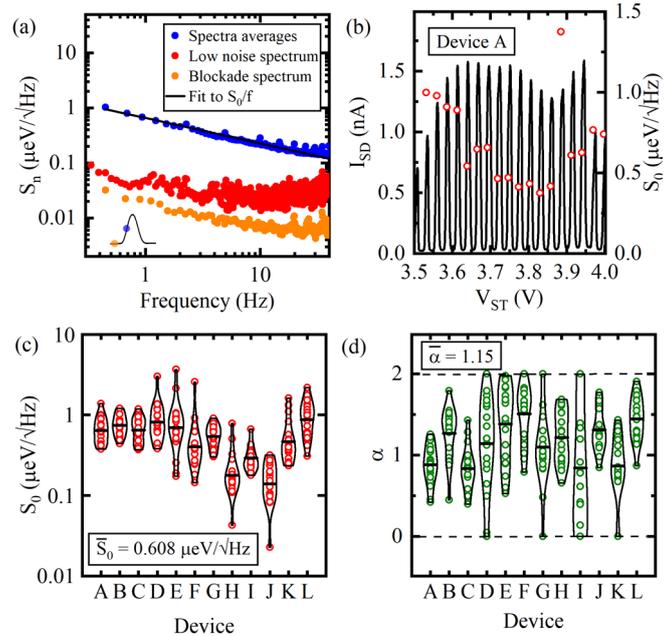

**Fig. 3| SET current spectroscopy and device statistics.** a) Noise spectral density of the quantum dot. The average spectrum of 231 measurements (blue) showing a standard $1/f$ spectrum. Spectral density of the charge noise in blockade (orange, low sensitive region) highlights that the background noise does not limit the measurements. b) Coulomb

oscillations of device A (black) by sweeping the ST bias. The charge noise at different peaks is overlayed. c, d) The violine plot of $S_0$ and $\alpha$ across 12 SETs and 231 peaks.

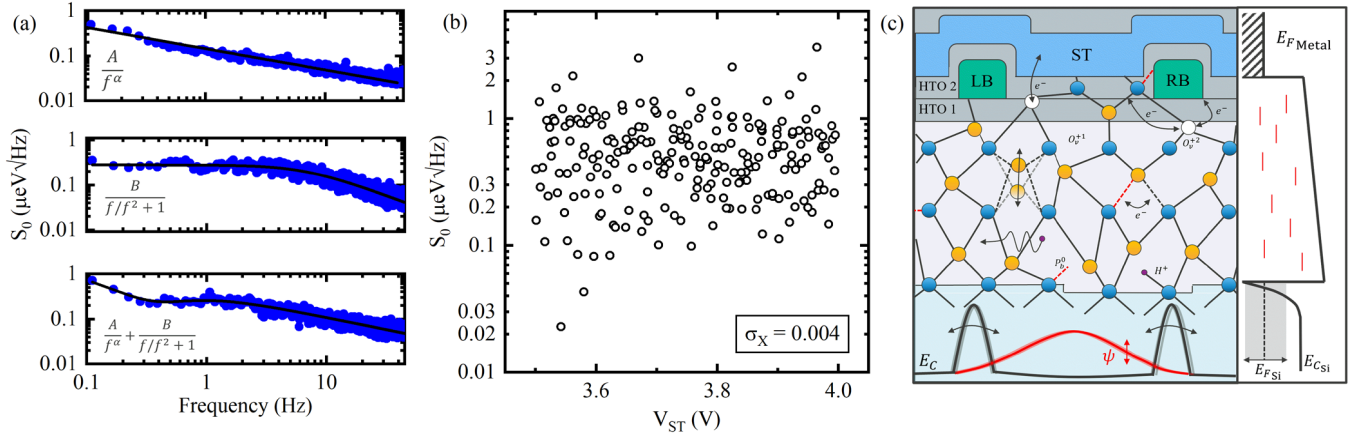

**Fig. 4| Charge noise metrics correlation.** a) Charge noise spectra with power law, Lorentzian, and combination of power law and Lorentzian. The Lorentzian spectrum can be observed when a single TLF dominates the charge noise spectrum. This is more likely to happen due to the random spatial distribution of TLFs and low overall TLF density. b) Scatter plot of $S_0$ with respect to $V_{ST}$. No correlation is observed suggesting negligible screening and uniform distribution of TLFs in energy. c) Schematic depicting the types of defects in SiOx (including dangling bonds, tunneling atoms, tunneling electrons and oxygen vacancies)[35]. The defects directly affect the electron wavefunction in the quantum dot. Right insert is the band diagram representation of the random TLF energy distribution in the oxide.

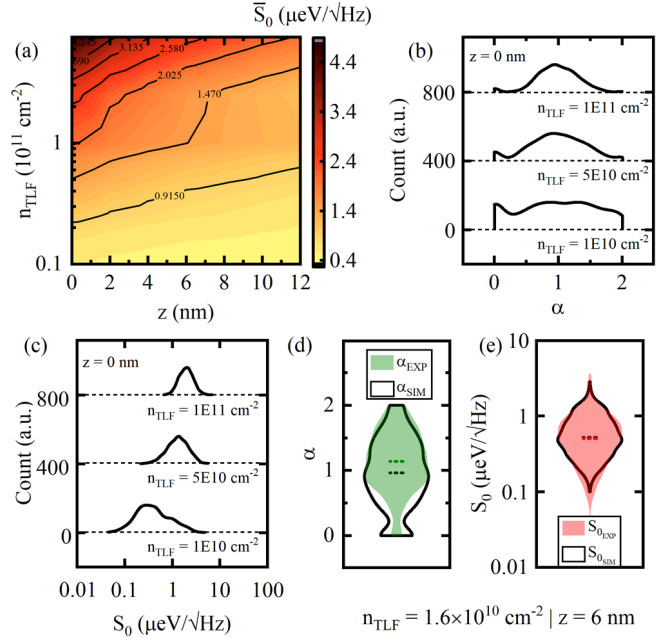

**Fig. 5| Charge noise simulations.** a) Contour plot of the simulated quantum dot charge noise with respect to the TLF density ($n_{TLF}$) and the depth in the oxide ($z$). b, c) The kernel distribution estimation of $\alpha$, $S_0$ at different $n_{TLF}$ generated from 1000 simulations. d, e) The violin plots of $\alpha$ and $S_0$ with experimental data and the best-fit simulation results ($n_{TLF} = 1 \times 10^{10}$ cm$^{-2}$, $z = 6$ nm).

# Supplementary information: Low charge noise quantum dots with industrial CMOS manufacturing

A. Elsayed[1,2], M.M.K. Shehata[1,2], C. Godfrin[1], S. Kubicek[1], S. Massar[1], Y. Canvel[1], J. Jussot[1], G. Simion[1], M. Mongillo[1], D. Wan[1], B. Govoreanu[1], I. P. Radu[1], R. Li*[1], P. Van Dorpe[1,2], K. De Greve[1,3]

[1] IMEC, Leuven, Belgium
[2] Department of Physics and Astronomy, KU Leuven, Leuven, Belgium
[3] Department of Electrical Engineering, KU Leuven, Leuven, Belgium

Contents

# 1. Mobility limiting mechanisms

Beyond the standard Hall mobility and percolation density, we further investigate the mobility limiting mechanisms. The electron transport scattering angle provides useful insight into the scattering center spacings: the further the scattering center to the electron channel, the smaller the angle. The relaxation time $\tau_t$ extracted from Hall mobility mainly reflects the large angle scattering. However, in a system where small angle scattering dominates, $\tau_t$ represents only a fraction of the actual scattering events. On the other hand, the quantum lifetime $\tau_q$ is a measure of the time for which an electronic momentum eigenstate can be defined even in the presence of scattering[1,2]. The quantum lifetime thus measures both small- and large-angle scattering which both equally dephase the cyclotron orbit. The quantum mobility, that follows from $\tau_q$, thus presents a more informative characteristic for qualifying the gate stack.

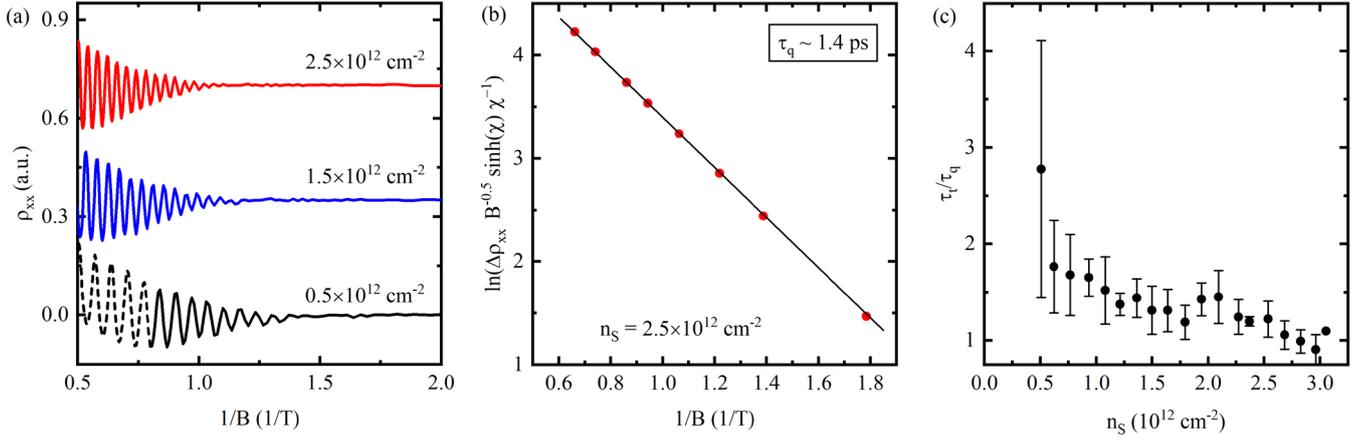

**Supplementary Fig. 1| Mobility limiting mechanisms.** a) Longitudinal resistivity $\rho_{xx}$ with respect to the applied magnetic field after subtracting the longitudinal magnetoresistance component. The plots are offset for clarity. b) The $\Delta\rho_{xx}$ envelope is then plot with respect to the inverse magnetic field. The data is fit to $\Delta\rho_{xx} \sim \sqrt{B}\chi_0/\sinh\chi_0 \, exp(-\pi/\omega_c\tau_q)$ to extract $\tau_q$. c) The Dingle ratio ($\tau_t/\tau_q$) for every density point measured. A Dingle ratio $\sim 3$ can be reached at lower electron density.

To extract the quantum lifetime, we investigate the Shubnikov DeHaas (SdH) oscillations of our 2DEG[1,2]. The envelope of the SdH oscillations is a function of the collision broadening of the Landau levels. Fig. S1a shows the low field oscillation amplitude ($\Delta\rho_{xx}$) with respect to applied field (B). The quantum lifetime is extracted from a fit to

$$\Delta\rho_{xx} \sim \sqrt{B} \cdot \frac{\chi_0}{\sinh(\chi_0)} \cdot \exp\left(-\frac{\pi}{eB\tau_q}\right)$$

with $\chi_0 = 2\pi k_B T_0/\hbar\omega_c$ where $k_B$ is the Boltzmann constant, $\hbar$ is the reduced Planck constant and $\omega_c$ is the cyclotron frequency. We extract $\tau_q$ for every density point and evaluate the Dingle ratio as seen in Fig. S1c. At higher densities the Dingle ratio is 1 as scatter centers at the interface are the dominating mobility limiting mechanism. However, the Dingle ratio $\sim 3$ is achieved at lower electron density, where the screening effect is weak, suggesting that defects inside the oxide also play an important role.

# 2. Extracting lever arm, dot features and slope

The lever arm is the ratio of the gate-to-dot capacitance to the overall dot capacitance[3]. To consistently extract the lever arm for the current to charge noise conversion and avoid calculation errors we employ an edge detection algorithm. We digitize the current to binary 1 (0) representing values above (below) the threshold current. Subsequently, we evaluate the first derivative which identifies the edges. At higher source drain bias the diamond edges typically round due cotunneling events. This rounding would give a larger uncertainty in the lever arm. Thus, we only consider edges in the low source drain bias range. We then plot the lines along the edges using the extracted slope, as shown in Fig. S2a. We find the points of intersection at zero source drain bias and at the corners of the diamonds using[3]

$$p = \frac{\begin{vmatrix} |a_1 \; a_2| & a_1 - a_2 \\ |b_1 \; b_2| & b_1 - b_2 \end{vmatrix}}{|a_1 - a_2 \quad b_1 - b_2|}$$

where the lines are $(a_1, a_2)$ and $(b_1, b_2)$. Finally, calculating the lever arm with $\alpha_L = \left((p_{1_L} - p_{1_R})/2\right)/(p_{c_1} - p_{c_0})$. Furthermore, we can determine the SET dot charging energies using $E_C = (p_{1_L} - p_{1_R})/2$.

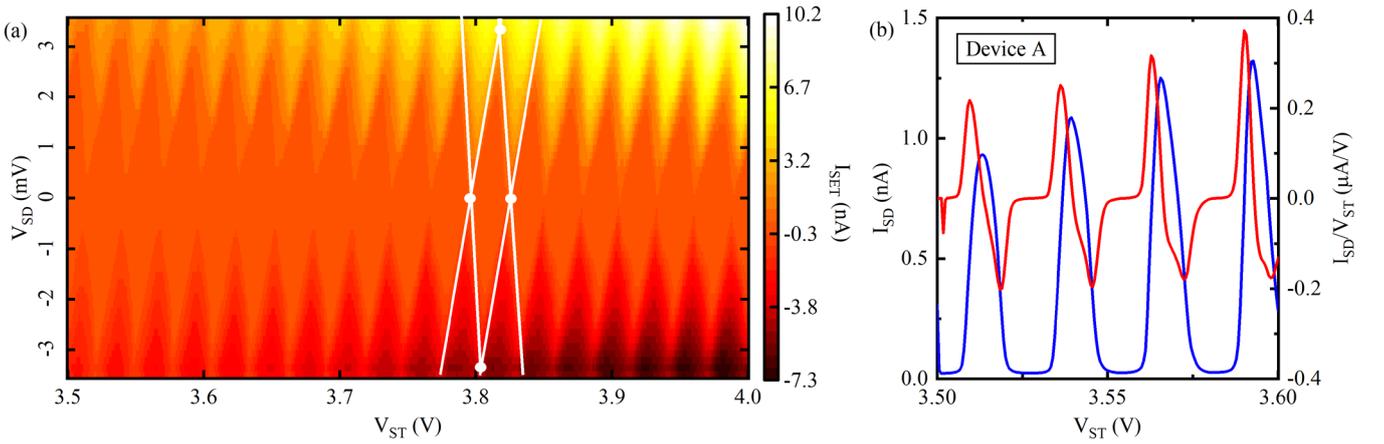

**Supplementary Fig. 2| Coulomb diamonds and dot features.** a) Coulomb diamonds of device A. The lever arm and charging energy is automatically extracted using an edge detection algorithm. b) The Coulomb oscillations at $V_{SD} = 1\ mV$.

The Coulomb oscillations are then measured at $V_{SD} = 1\ mV$ and the slope $dI_{SD}/dV_{ST}$ is used to convert the current noise to a voltage noise as shown in Fig. S2b (also see methods).

## 3. Noise spectra fitting and charge noise correlations

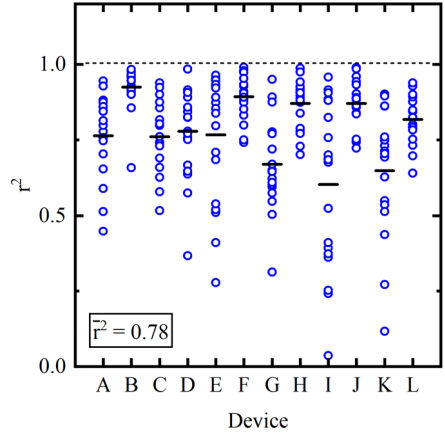

**Supplementary Fig. 3| Noise spectra fitting.** The residuals are calculated and used to determine the best fit for each spectrum measured. The device average is denoted with a dash.

To have consistent analysis of the 231 measured noise spectra, we automate the procedure as follows. First, we convert current noise to charge noise using the respective $dI/dV_{ST}$ and lever arm $\alpha_L$. By default, we fit the charge noise spectrum to $S_0/f^\alpha$ then calculate the residuals $r^2$ value to determine the reliability of the fit. Below a certain threshold the fit is flagged and attempted with a Lorentzian or a sum of a power law and Lorentzian. Finally, any fits that are below the quality threshold are inspected individually to see which dependence best describes the spectrum. We report an overall $r^2 = 0.78$ and any low $r^2$ can be attributed to a larger distribution around the power or Lorentzian spectrum (mainly in the low noise spectra due to other noise functions).

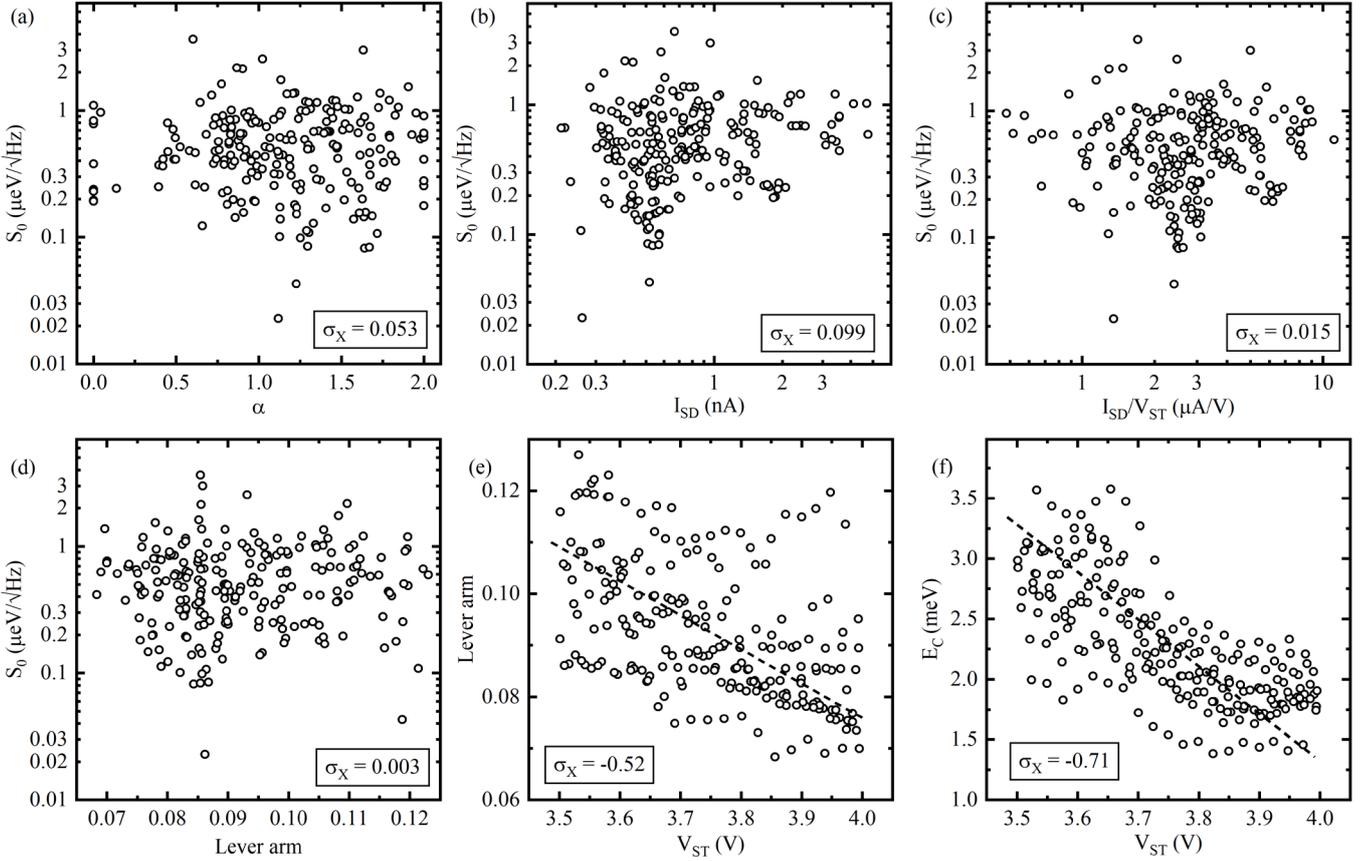

**Supplementary Fig. 4| Noise spectra fitting and noise correlations.** a) Scatter plot of $S_0$ with respect to the fitting exponent $\alpha$ showing a covariance of ~ 0 suggesting that the TLFs are uniformly distributed in frequency. b, c) The scatter plots of the $S_0$ with respect to $I_{SD}$ and $I_{SD}/V_{ST}$ at the measurement point, showing no covariance. d) The scatter plot of $S_0$ with respect to the lever arm showing no covariance suggesting that the size of the dot does not affect $S_0$. e, f) The scatter plot of the lever arm and the charging energy with respect to the top gate voltage respectively.

In Fig. S4 we show the correlations of charge noise with respect to different metrics. We calculate the covariance ($\sigma_X$) of different key metrics with respect to $S_0$ for further insight to the origin of the noise. In Fig. S4a we observe no correlation between $S_0$ and $\alpha$. In Fig. S4b, c we observe no covariance between $S_0$ and $I_{SD}$ at the measuring point nor between $S_0$ and $I_{SD}/V_{ST}$ of the measurement point. This suggests that the shot noise, which originates from the stochastic electron tunneling events and is typically proportional to the current magnitude, is not the limiting factor.

As the top gate voltage is increased, the size of the dot changes and the capacitance ratio to the top and barrier gate also changes. This is evident in the negative correlations seen between the lever arm and the charging energy $E_C$ with respect to $V_{ST}$ in Fig. S4e and f respectively. This negative correlation coupled with the lack of correlation between the lever arm and $S_0$ (Fig. S4d) further confirm that the screening is negligible. In addition, above plots also affirm the procedure of extracting the noise spectra from the SET current fluctuations, as $S_0$ is independent of the shape and amplitude of Coulomb peaks.

## 4. Slow drift noise

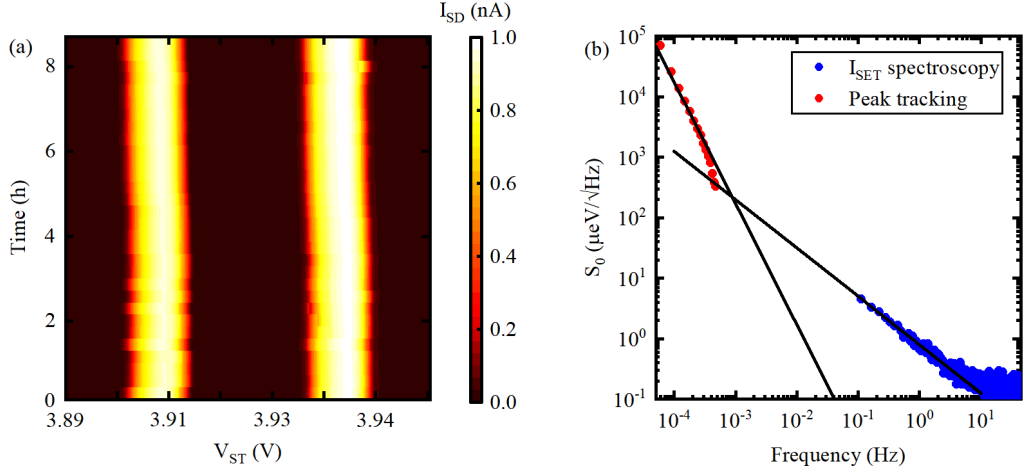

**Supplementary Fig. 5| Slow drift noise.** a) Coulomb oscillations with respect to time showing very small drift over 8 hours. b) The resulting charge noise spectrum of the peak tracking and the noise from the SET current at the same Coulomb peak.

The long-term drift was measured across two peaks on two different devices (F and I). We record the full Coulomb peaks over the duration of 8 hours, as seen in Fig. S5a. The peaks are then fit using a combination of Gaussian and Lorentzian distributions (attributed to both tunnel and thermal broadening). The peak drift is then converted to the frequency domain and plotted together with the charge noise from the SET current, as shown in Fig. S3b. The slow noise spectrum is fit using the same methods discussed previously. We find that the four peaks measured have an $\alpha = 2$ which has the signature of Brown noise[4]. The monotonic drift of the Coulomb peak strongly resembles the motion of mobile charges in the gate stack[4], which could be fundamentally different from the mechanisms causing the charge noise around 1 Hz. Though it is hard to determine the exact origin of the drift, the extrapolated noise level at 1 Hz is very low, and the overall drift is very small. It will not be a limiting factor for qubit operations and can be corrected in feedback or retuning schemes.

## 5. SET charge noise during QD operation

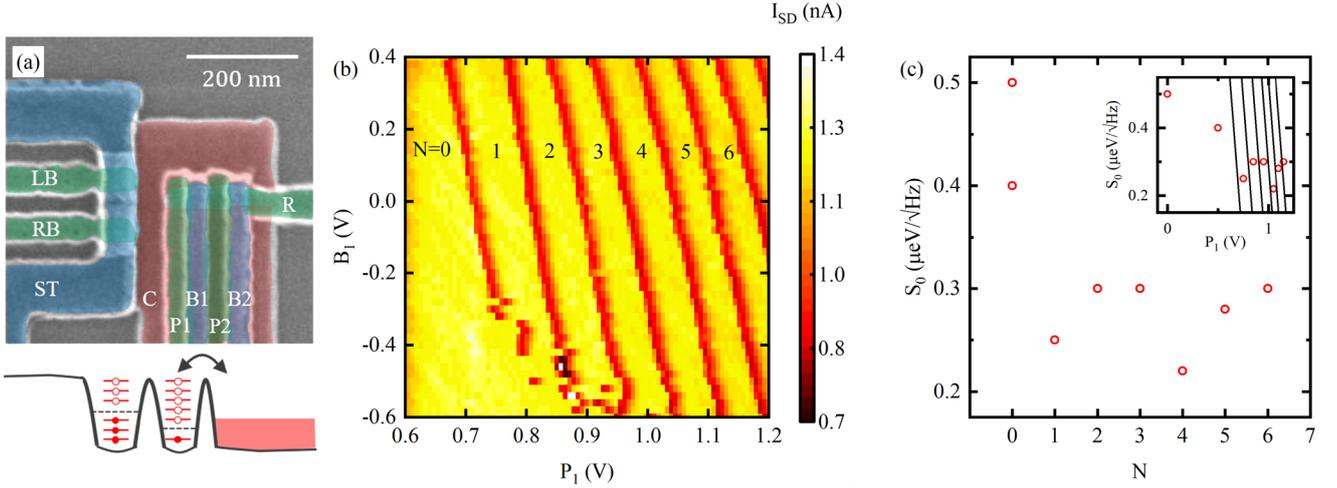

**Supplementary Fig. 6| SET charge noise during QD operation.** a) SEM micrograph of the SET neighboring a double QD structure. Only dot 1 was tuned and the SET subsequently measured. b) Charge stability map of QD1 showing controlled filling down to the last electron. c) Charge noise with respect to the dot occupation (N).

We characterize the charge noise of the SET during normal operation of the neighbouring QDs (Fig. S6a). At a fixed SET operating point, we characterize the dot under P1 and show controlled electron filling down to the last electron (Fig. S6b). We measure the charge noise on the SET for each P1 electron occupation ($N$). The charge noise with respect to $N$ is plotted in Fig. S6c. We find no clear correlation between the SET noise and the dot occupation.

## 6. Charge noise simulations

To simulate the charge noise, we consider double well potential systems (DWPs) type of TLFs, which can be structural defects, atomic vacancies, and other types of microscopic defects. As shown in Ref 5, DWP best represent the charge noise measured in quantum systems. To model the effect of DWPs, we use dipoles with a length of 1 Å oriented horizontally in the $xy$-plane with switching frequencies distributed log-unform over $10^{-2}$ Hz to $10^6$ Hz. Subsequently, the total spectral density can be calculated from:

$$S_N = \sqrt{\sum_i^{N_t} \frac{\Delta\mu_i^2 \tau_i}{4 + (2\pi f)^2 \tau_i^2}}$$

where $N_t$ is the total number of DWPs, $\tau_i$ is the characteristic switching time of each DWP and $\Delta\mu$ is the magnitude of the electrostatic fluctuation caused by the DWPs which is calculated perturbatively as described in ref 5. A single simulation involves creating 1000 different spatial distributions of the horizontal DWPs distributed in a 2D plane situated in the oxide. The $S_0$ and $\alpha$ are determined for each of the 1000 random spatial distributions using the same procedure described for the experimental results. This full procedure is repeated for varying the density ($n_{TLF}$) and the depth ($z$) as discussed in the main text.

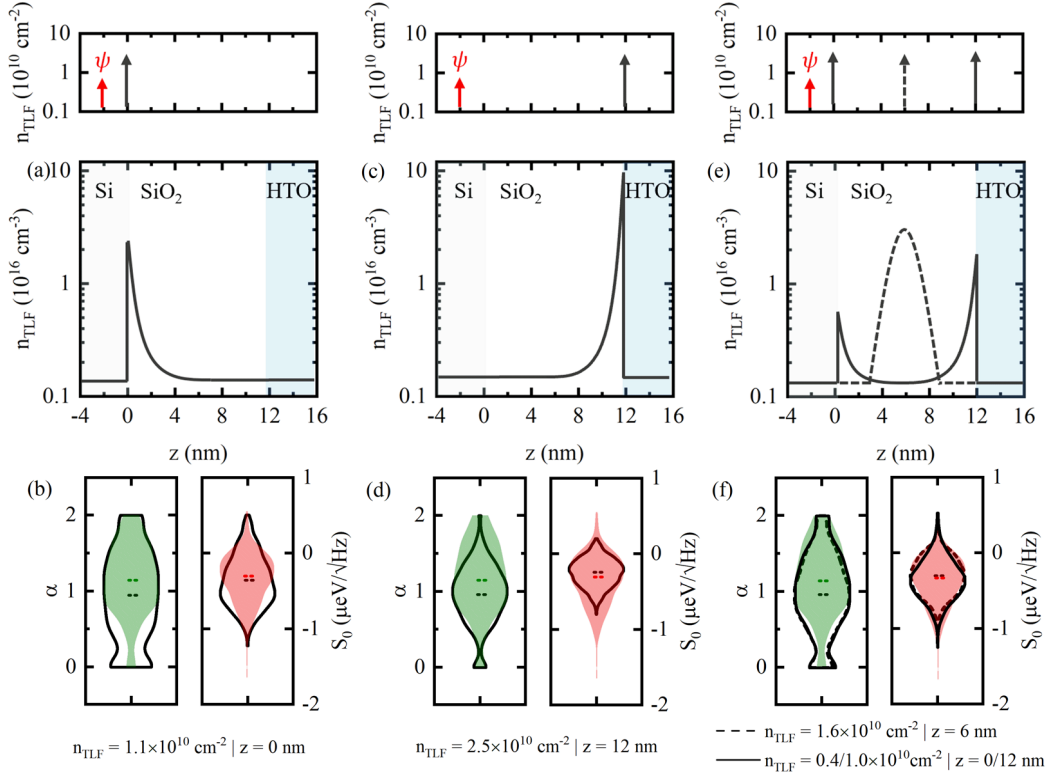

**Supplementary Fig. 7| Charge noise simulations.** (a, c, e) The defect density volume distribution in the oxide and the effective 2D defect density used in the simulation. The electron wavefunction is assumed to be $2\ nm$ below the oxide. (b, d, f) The KDE of $\alpha$ and $S_0$ extracted from the model (black line) using the 2D effective density with the experimental KDE overlayed (color fill).

In Fig. S7, we show the histograms of the simulated $S_0$ and $\alpha$ overlayed on the experimental histograms with different $n_{TLF}$ and $z$. We find $n_{TLF} = 1.6 \times 10^{10}\ \text{cm}^{-2}$ at $z = 6$ nm fits very well to the experimental distributions. $n_{TLF} = 1.1 \times 10^{10}\ \text{cm}^{-2}$ at $z = 0$ nm and $n_{TLF} = 2.5 \times 10^{10}\ \text{cm}^{-2}$ at $z = 12$ nm could still match the experimental distributions to some extent. Interestingly, we find that simulating the charge noise using two 2D defect density distributions at both the semiconductor/oxide interface and the upper oxide interface also fits the experimental distributions very well. In our previous work[6,7], we studied the Hall mobility of the Si MOS interface with 8 nm oxide, which, in comparison to the 12 nm oxide in this study, gives lower peak mobility, higher percolation, and a dingle ratio closer to 1. We argue that the improved Si/SiOx interface quality with increased oxide thickness and the better matching of the simulation result with TLFs into the oxide strongly suggest defects in the oxide play an important role and the Si/SiOx interface does not need to be detrimental for spin qubit upscaling.

## 7. Two qubit gate fidelity

To highlight the importance of the charge noise environment on the qubit operation we feed three of the regenerated random TLF distributions into a numerical model to evaluate the SWAP gate fidelity. The three distributions are chosen to have low, average, and high $S_0$ (0.23 µeV, 4.05 µeV and 0.6 µeV). The SWAP gate is driven by exchange which has an exponential dependency on the electrical potential. We use it to highlight the effect of charge noise on qubit operations. The Hamiltonian driving the SWAP gate can be described as:

$$H(t) = \frac{1}{2}\begin{pmatrix} 0 & J + \Delta J(t) \\ J + \Delta J(t) & 0 \end{pmatrix}\begin{pmatrix} |\uparrow,\downarrow\rangle \\ |\downarrow,\uparrow\rangle \end{pmatrix}$$

where $J$ is the unperturbed exchange interaction, and $\Delta J$ is the shift in exchange due to the TLFs. Without noise and an initial state of $|\downarrow,\uparrow\rangle$, the Hamiltonian above drives the SWAP operation where the probability of the final $|\uparrow,\downarrow\rangle$ state is described by:

$$P_{\uparrow,\downarrow}(t) = sin^2\left[\frac{J \cdot t}{2}\right]$$

Using the double quantum dot (DQD) model in ref 5 the shift in the exchange interaction due to the TLFs is calculated. The average SWAP gate fidelity can be calculated using:

$$F_{SWAP} = \frac{1}{N_{SWAP}} \Sigma_n^{N_{SWAP}} sin^2\left[\frac{(J + \Delta J(t_n))t_S}{2}\right]$$

where $t_S$ is the SWAP gate time in the ideal system ($\Delta J = 0$). The shift in exchange ($\Delta J$) is sampled $N_{SWAP}$=1000 times with $t_n - t_{n-1} \approx t_S$ to simulate 1000 SWAP gate operations. The final SWAP errors are shown in Fig. S8, giving an average gate fidelity of 99.98% with the low noise distribution and 99.75% for the high noise distribution.

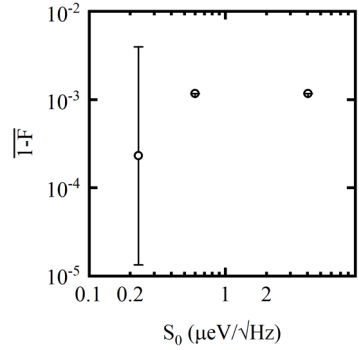

**Supplementary Fig. 8| SWAP gate fidelity.** Average SWAP gate infidelity for the lowest, average, and highest simulated $S_0$.